\documentclass[a4paper,12pt]{article}
\usepackage{authblk}
\usepackage[colorlinks,allcolors=blue]{hyperref}
\usepackage[T1]{fontenc}
\usepackage[utf8]{inputenc}
\usepackage{amsmath}
\usepackage{setspace}
\usepackage{graphics}
\usepackage{graphicx}
\usepackage{geometry}
\newgeometry{margin=2.5cm}
\begin{document}
\title{Collective excitation of bosonic quantum Hall state}
\author{Moumita Indra\footnote{Orcid ID: 0000-0002-7900-7947 \\ All the correspondence should be addressed to the e-mail: M. Indra (moumitaindra@ee.iitb.ac.in / moumita.indra93@gmail.com)} \;  \& \; Sandip Mondal}
\affil{Department of Electrical Engineering, Indian Institute of Technology Bombay, Mumbai 400076, India}
\date{}

\maketitle
\doublespacing
%
\vspace{0.5cm}
\begin{abstract}
The recent developments in the theory of rapidly rotating Bose atoms have been reviewed in this article. Rotation leads to the development of quantized-vortices, that cluster into a vortex array, exactly to how superfluid helium behaves. Theoretically, a number of strongly correlated phases are projected to exist in this domain, which might be thought of as bosonic counterparts of fractional quantum Hall effect (FQHE). It is now possible for bosons associating with a short-range interaction to exhibit a FQHE, because the system of neutral bosons in a fast rotating atomic trap is analogous to charged bosons placed in a fictitious magnetic field. The neutral collective spin-conserving and spin-flip excitation for the rotating ultra-cold dilute Bose atoms in the FQHE domain are being discussed.
We have introduced a realistic interaction between the Bose particles together with long-range interaction and presented a short review article about various fractional quantum Hall states and their spin conserving and spin reversed collective modes.
\end{abstract}
\pagebreak
\tableofcontents
\pagebreak
\section{Introduction}
Bose-Einstein condensation (BEC) of atomic gases was first experimentally realized in 1995 \cite{Davis_1995, Ketterle}, and it has since led to a wide range of phenomena.
 Atomic BEC can produce vortex lattices by being confined in a magnetic trap and being stirred with a rotating deformation called vortex \cite{BEC_vortex, BEC_vortex1}. When the number of vortices equals the number of atoms, one might imagine entering a phase of even greater vortex density \cite{Hamiltonian}. In trapped rotating BEC, a fascinating Abrikosov vortex lattice (triangular lattice) emerges, as described by various experimental groups including the MIT group \cite{Davis_1995} in 1995, the JILA group \cite{GP_equn, GP_equn1} in 1999, the ENS group \cite{vortex} in 2000, and others. Several researchers have also created this vortex lattice theoretically \cite{theo_vortex, theo_vortex1}. Another interesting study by Saswata {\it et. al.} \cite{Saswata} showed that in the case of non-uniform rotation, vortices are formed in a circular ring shape above a specific amount of rotation. The emergence of highly correlated phases in this situation is predicted by theoretical research; that is why rotating BEC becomes a hot topic of research nowadays. Rotating BEC trapped in a harmonic potential is analogous to the fractional quantum Hall states (FQHS) of electrons in semiconductors \cite{Girvin}. Several intriguing and unique characteristics such as fractional plateaus, and dissipation-less current flow are observed in FQHS and these states are topologically protected. The theoretical knowledge of rotating Bose atoms in these unique regimes has advanced recently, as have experimental techniques. This article's goal is to evaluate recent theoretical improvements. 
  Even though researchers still don't fully comprehend these systems, theoretical research has helped to clarify some of their anticipated characteristics. These theoretical findings are expected to inspire and guide future experimental studies of these unique occurrences.
\section{A brief description of the quantum Hall effect}
One of the most striking and surprising achievements in condensed matter physics is the quantum Hall effect. After the scientific breakthrough of the integer Quantum Hall Effect (IQHE) in 1980 \cite{IQHE}, and fractional quantum Hall effect (FQHE) in 1982 \cite{fqhe, Stormer_RMP99}, two-dimensional electron systems (2DES) subjected to a perpendicular magnetic field remain a very active field of research experimentally or on the theory level throughout the last four decades. Hall resistivity shows a strange step-wise variation with the magnetic field, as an integer or fractional multiple of $h/e^2$ (where $h$ be the Planck's constant and $e$ be the charge of an electron) and the longitudinal resistivity also displays severe minima at that location. FQHE is basically a collective phenomenon, that occurs in a strongly interacting system. In many-body systems, quasi-particles are typically employed to acquire the system's ground and excited states as well as to calculate the overall excitation and estimate the bulk characteristics of low-energy systems. Electrons produce new quasi-particles (collective states) in FQHS, that obey anyonic statistics and fractional charge \cite{QAHE}. 
 \begin{figure}[h]
 \centering
   \includegraphics[width=12cm]{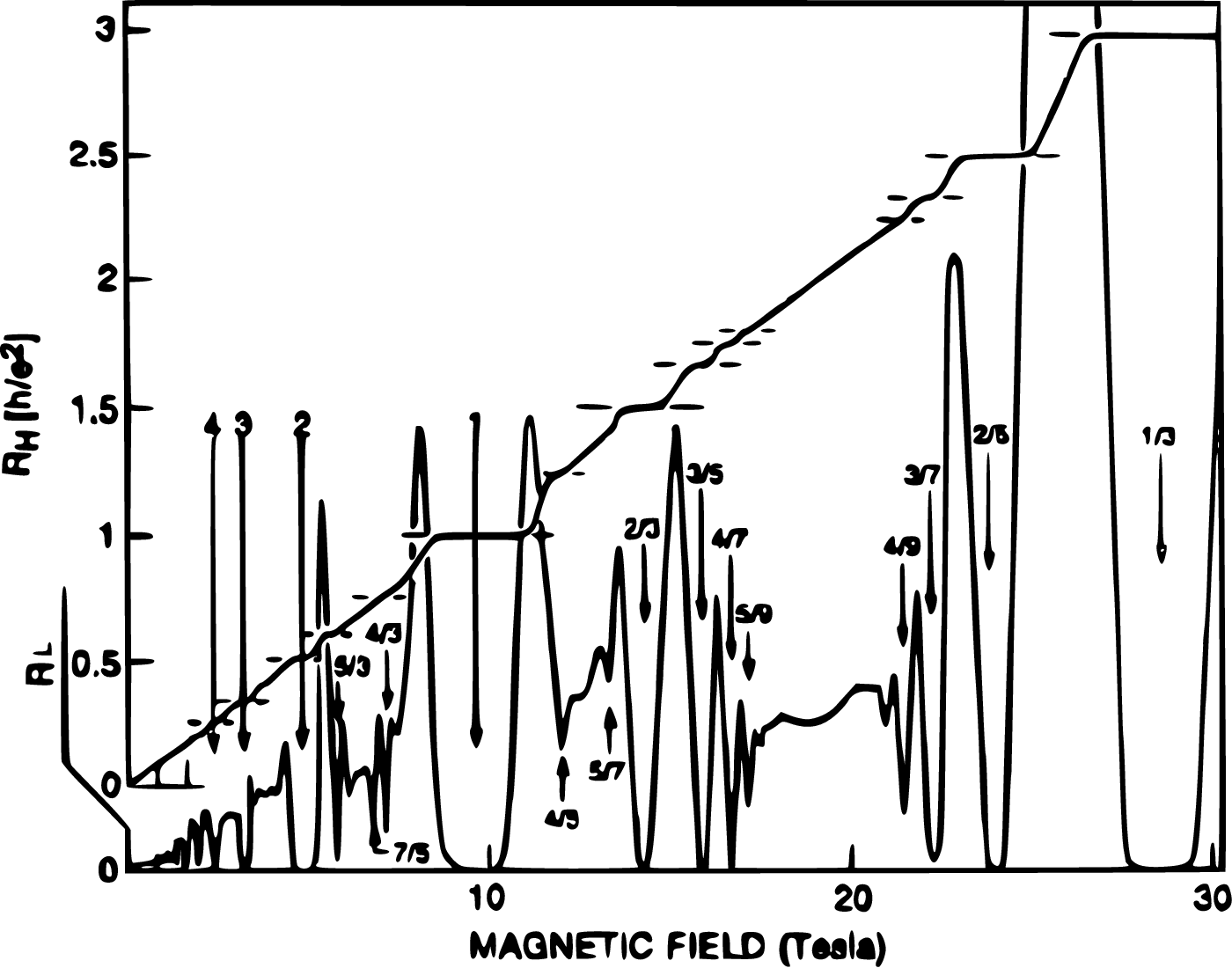}
   \caption{Experimental data of fractional plateaus in ultra-cold temperature, high mobility modulation-doped 2DESs: The step-wise behavior of the transverse resistance $R_H$, superimposed with the longitudinal resistance $R_L$, as a function of the magnetic field. Many fractions are observed, the most prominent sequence is $\nu = \frac{n}{2n \pm 1}$ with integer values of $n$ (From Ref \cite{Stormer})}.
   \label{FQHE}
 \end{figure}
\section{The Jain sequence using composite fermions}
Composite fermion (CF) theory \cite{Jain_CF} is proposed by J. K. Jain to explain the fractional filling fraction. According to this theory, it is energetically desirable for the electrons in the lowest Landau level (LLL) to pull in an even number ($2f$) of vortices from the externally applied magnetic field ($B$). Electrons with attached vortices behave like a single particle, called the quasi-particle (CF). Aharonov-Bohm phases produced from the external magnetic field is partially canceled by Berry phases created by the vortices attached with the electrons; since the CFs are moving around.
\begin{figure}[htbb]
 \centering
   \includegraphics[width=0.45\textwidth,angle=0]{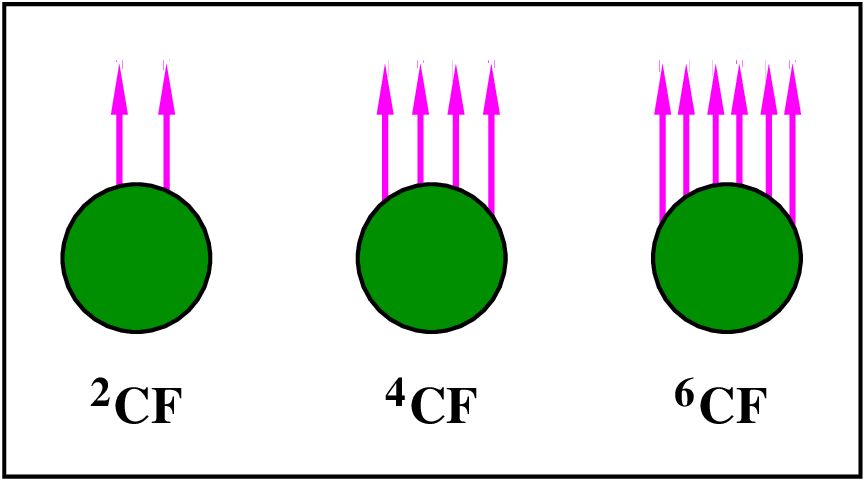}
   \caption{Pictorial representation of three flavors of composite fermion carrying two, four, and six vortices respectively (Ref. \cite{Jain_Book}).}
 \end{figure}
Thus, an effective weaker magnetic field is experienced by the CFs. CF can detect the following  magnetic field
 \begin{equation}
 B^* = B - 2 f \rho \phi_0 \;;
 \end{equation}
where the electron density is denoted by $\rho$ and $\phi_0 = \frac{hc}{e}$ be the flux-quanta.
The Landau levels (LLs) with reduced degeneracy are analogous to the eigenstates of the Hamiltonian of a CF, which are referred to as $\Lambda$-levels. The FQHS of filling fraction $\nu$ of a strongly interacting electronic system is described by $n$ number of filled $\Lambda$-levels of non-interacting CFs (integer quantum Hall effect of CF). As a result, both filling fractions are related by a sequence known as Jain's sequence
\begin{equation}
 \nu= \frac{n}{2fn + 1} \;.
 \label{CF_seq}
\end{equation} 
In the case of negative flux attachment, the electron adds up some vortices to the external magnetic field. Then the CF perceives an enhancement in the effective magnetic field as well. Then the relation (\ref{CF_seq}) is modified as,
\begin{equation}
 \nu= \frac{n}{2fn - 1} \;.
\end{equation} 
\section{Bose-Einstein Condensate---a brief overview}
Bose-Einstein condensate (BEC) is a state of matter that occurs at nearly absolute zero temperature when all the atoms of the system occupy the lowest ground state. The atoms gradually come to the lower energy states during the cooling process, and their wavelength starts to overlap. Finally, below a critical temperature ($T_c$) near absolute zero, the majority of the atoms occupy the
lowest energy state, and atoms collectively form a single giant matter-wave. One of the fundamental conditions to achieve BEC is that the system must be ultra-dilute. As an example the solid has a density of $10^{38} cm^{-3}$, whereas in case of liquid, it is $10^{22} cm^{-3}$ and in the case of gas, the density is $10^{15} - 10^{19} cm^{-3}$. For BEC the density is typically $10^{13} - 10^{15} cm^{-3}$. After seventy years of this theoretical prediction of BEC by Albert Einstein following S. N. Bose, in 1995, the first it has been achieved experimentally in alkali atoms. It was Ketterle \cite{Davis_1995} and Cornell \cite{BEC} who first created the BEC of magnetically confined alkali atoms. Following this discovery, a quantum state of matter was produced by chilling a low-density bosonic gas to nearly absolute zero degree temperature, where quantum mechanical phenomena (wavefunction interference) may be observed. The BEC has the advantage of magnifying tiny quantum phenomena to macroscopic sizes that may be seen; that's why condensates have attracted attention. More complicated particles may now be condensed in both the fermionic and bosonic domains, including lanthanides \cite{Lanthanide}, reactive metals linked to high-temperature superconducting \cite{supercond}, and quantum droplets made of Bose-Bose mixtures and dipolar gases (supersolids) \cite{supersolid}. 
\begin{figure}
 \centering
   \includegraphics[width=10cm]{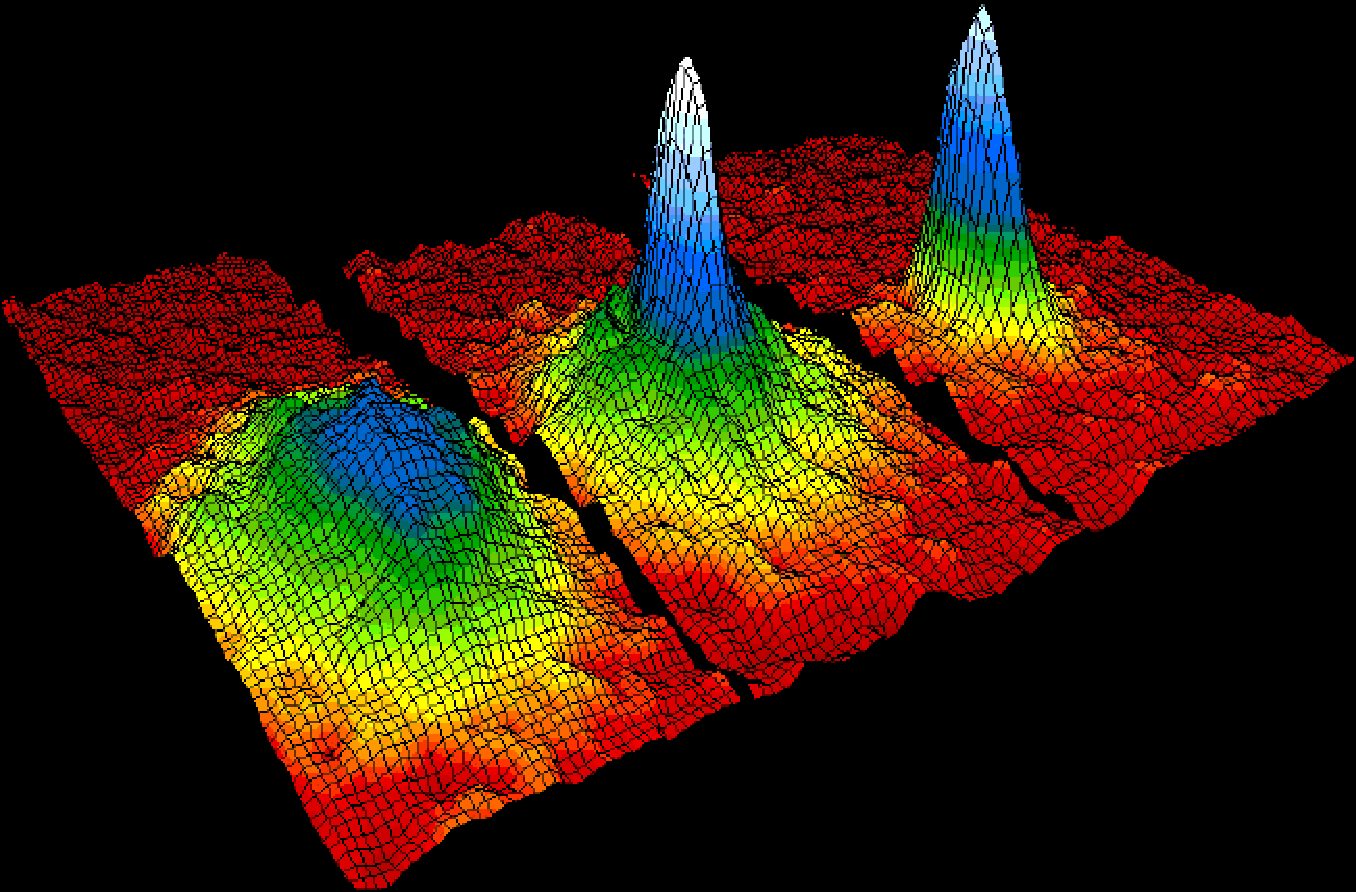}
   \caption{Three views show velocity distribution data for a gas of rubidium atoms, confirming the identification of the Bose-Einstein condensate. The gas shifted from a broad, active thermal gas to a confined distribution (left to right), which is typical of the condensate, as the temperature dropped (Data taken from: NIST/JILA/CU-Boulder).}
   \label{Vel_BEC}
 \end{figure}
%
\section{Dilute rotating Bose atoms at ultra-cold temperature}
Rotating BEC has recently been shown to provide an intriguing platform for the investigation of FQHE \cite{Chang}. First Cooper and Wilkin \cite{CF_on_BEC} applied a non-interacting composite fermion (CF) model to predict the first FQHE in the ultra-cold rotating 2D Bose atomic system. A hypothetical magnetic field is produced (along Z-axis) perpendicular to the rotating Bose atoms confined in a 2D-harmonic trap (say, XY-plane); which is equivalent to the 2DES in the presence of a perpendicular magnetic field. It results in the formation of Landau levels (LLs). Then, there is a prospect that the FQHE will occur in a fast-rotating BEC of dilute neutral Bose gas \cite{BEC_rotation}. The majority of the particles fly away when a high rotation is imparted to a bosonic system, which results in an extremely low atomic density. When the temperature and interactions are sufficiently low in comparison to the LLs' cyclotron gaps, bosons could entirely dwell within the LLL. Since the kinetic energy is quenched, this issue can only be accurately determined by the interactions between the particles. Through the investigation of emergent collective behaviour, also known as collective excitations, quantum matter sets in a new era in the study of many-body quantum systems. In this low-density and theoretically absolute zero temperature, the inter-particle spacing and de-Broglie wavelength are much larger than the scattering length. So, the interaction between the two atoms of the system can be expressed as the delta-function like contact interaction. Therefore, at low energy only binary collisions are significant in such a system, and these collisions are described by the s-wave scattering length only \cite{s_wave}.
 Regnault and Jolicoeur used exact diagonalization to study the ground state and low-lying excited states \cite{exact_12} of FQHE in the diluted limit of rotating BEC with a few numbers of bosons. For the filling fraction $\nu=1/2$, it has been observed that the CF theory and the exact diagonalization result match quite well \cite{Chang}. Additionally, it is confirmed by Moumita {\it et. al.} \cite{SDE_MI, MI23} that the CF wavefunction and the exact diagonalization coincide well for the filling fractions of  $1/4$ and $1/6$. Orion Ciftja studied Bose-Laughlin wave function \cite{Ciftja} for filling factors 1/2, 1/4 and 1/6 and reported that Bose-Laughlin ground state energy is practically identical to the energy of the true CF liquid state. Magnetic hyperfine states of the same atomic species can be trapped, as well as mixtures of two separate atomic species. Both of these processes result in the formation of two-component BECs. By using sympathetic cooling, it is possible to combine two condensates that correspond to the two separate spin states of $^{87} Rb$; $ \mid F=1, \; m=-1 >  $ and $ \mid F=1, \; m=0 >$ \cite{2BEC}. The hyperfine states of the two-component sodium ($^{23}{Na}$) condensate i.e. $\mid F=1, \; m=1,0>$ were also predicted in experiments \cite{PRL'82}. They generated magnetic domain-based long-lived metastable excited states. The experiment also includes the realization of a combination of BECs for the atomic species $^{41}K$, $^{87}Rb$ \cite{PRL'89, PRL'100} and $^{85}Rb$, $^{87}Rb$ \cite{PRL'101}. The two-component system implies that the spin-polarized states and the more complicated structure of FQHE may exist \cite{2Comp_BEC} and can be produced by the sympathetic cooling process \cite{Sympath_cooling}. Rather than the fully spin-polarized state, we've taken into account spin-conserving and spin-flip excitation of spin-polarized FQHS of Bose atoms. Throughout this work, just two-hyperfine states are considered.
\subsection{Hamiltonian of the system}
The Hamiltonian for $N$ interacting Bose atoms of mass $m$, rotating with angular velocity $\Omega\hat{z}$ confined in a 2D isotropic harmonic trap (XY-plane) of frequency $\omega$ can be expressed as
\begin{eqnarray}
H &=&  \sum_{i=1}^N \left ( \frac{p_i^2}{2m}+\frac{1}{2} m \omega^2 r_i^2 \right ) - \Omega L_z + \sum V(r_{ij}) \;,
\end{eqnarray}
where $\vec{r_i}$, $\vec{p_i} $ is the position, momentum of i-th atom respectively.
The z-component of angular momentum is,
 \begin{eqnarray}
 L_z &=& \sum_i \left ( x_i p_{iy} - y_i p_{ix} \right ) \;. \nonumber
 \end{eqnarray}  
$V$ is the interaction potential between the atoms.
The Hamiltonian can be rewritten as
\begin{equation}
H = \sum_{i=1}^N \frac{1}{2m} \left (\vec{p}_i-\vec{A}\right )^2 + \left ( \omega-\Omega \right ) L_z + \sum V(r_{ij}) \;,
\label{H_Bose}
\end{equation}
where the vector potential is chosen as
\begin{eqnarray}
  \vec{A} &=& m\omega \left (-y, x \right )  \;.
\end{eqnarray}  
The above Hamiltonian (equation \ref{H_Bose}) is equivalent to the Hamiltonian of a system of charged particles in presence of magnetic field, with vector potential $\vec A$ \cite{Hamiltonian}. The kinetic energy of the single-particle Hamiltonian become quantized with $E_n=\left (n+\frac{1}{2} \right ) \hbar \omega_c$ where $\omega_c = 2 \omega$. We assume that the system is sufficiently dilute so that all the atoms are confined in the LLL, the Hamiltonian becomes
\begin{eqnarray}
  H = \left (\omega-\Omega \right ) L_z + V  \;.
\end{eqnarray}
Here we can assume $\omega = \Omega$. So the problem becomes identical with the FQHE in 2DES.
\subsection{Interacting potential between dilute cold Bose atoms}
\subsubsection{\bf{Short ranged contact interaction}}
\begin{figure}
 \centering
   \includegraphics[width=12cm]{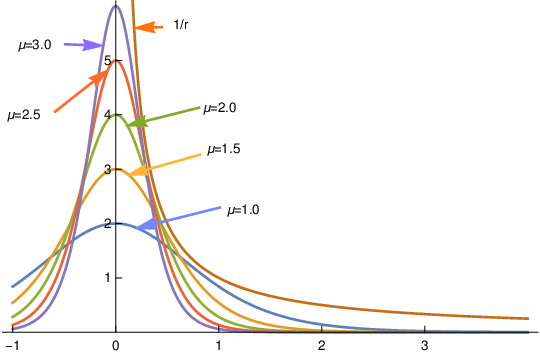}   \caption{For various values of $\mu$ (having unit inverse of $l_B$), the P$\ddot{o}$schl-Teller interaction potential, $V_{PT}$, is plotted as a function of the interparticle separation distance in the unit of magnetic length ($l_B$) [ Data taken from Ref. \cite{Physica_B} ].}
   \label{deltapotl}
 \end{figure}

Theoreticians used the concept to project interacting Bose particles onto weakly interacting spin-less fermions \cite{QH_spinlessBOSON, Jain_2006}, and they also took into account the $\delta$-function type interaction \cite{Delta_pot} between the Bose atoms. Only for ultra-cold diluted bosons the scattering between the atoms take place in the s-wave.
At low energy limit the effective interactions between the bosons can be expressed by a constant term in momentum space $U_m=4\pi \hbar^2 a_s/m$ \cite{exact_12}, where $m$ denotes the mass of Bose atoms and $a_s$ be the s-wave scattering length. So the Fourier transformation of this interaction in the coordinate space is the $\delta$-function type, contact interaction. 
The interaction in real space may be formulated as
\begin{eqnarray}
V = U_m \sum_{i<j} \delta^{(3)} \left ( \vec{r}_i - \vec{r}_j \right ) \;.
\end{eqnarray}
In this interaction, the condensate is governed by a non-linear form of Schr$\ddot{o}$dinger equation, which is known as Gross–Pitaevskii equation, first investigated in the superfluid system \cite{superfluid}. It is common practice to solve that equation numerically to study the different properties of the condensate \cite{PRA56_1996, JCP_2003} in the field of rotating BEC. We consider the system with strong confinement along Z-axis so that we can treat the system as 2D. So the interacting potential is
\begin{eqnarray}
V = U \sum_{i<j} \delta^{(2)} \left ( \vec{r}_i - \vec{r}_j \right ) \;.
\end{eqnarray}
{\bf{\textbf{P$\ddot{o}$schl-Teller interaction:}}}
In quantum Monte Carlo calculations, the $\delta$-function potential is extremely challenging to manage, and it also consumes a lot of computer power. It is not possible to determine the energy spectra of large particles using $\delta$-function potential. We have studied the P$\ddot{o}$schl-Teller interaction potential ($V_{PT}$) \cite{Poschl-Teller, PT_int} to get over this issue and access the system size in the thermodynamic limit. It has a form
\begin{eqnarray}
V_{PT} = U \sum_{i<j} \frac{2\mu}{cosh^2 \;( \mu  \; r_{ij})} \;,
\end{eqnarray}
where 1/$\mu$ stands for the interaction's width and $\mu$ denotes a parameter of interaction. By changing the value of $\mu$, we can vary the range of interaction. In order to explore the interaction range dependent nature, a variety of $\mu$-values have been taken into consideration. The nature of potential changes to a $\delta$-type when we increase the value of $\mu$, as seen in Fig. \ref{deltapotl}. The $1/r$ plot (figure's outermost line) is also displayed to compare the PT-interaction with that of Coulomb.
\subsubsection{Long-range Coulomb potential}
We are especially thrilled to witness the collective excitation in FQHE in the long-range $1/r$ interaction too since FQHE happens in the 2DES under the strong repulsive Coulomb potential, even though such potential is hardly ever seen in cold atoms.
\begin{eqnarray}
V_{Cou} = \sum_{i<j} \frac{C}{ r_{ij}} \;,
\end{eqnarray}
where $r_{ij}$ is the separation between two particles and $C$ is the coefficient, which is made up of all the data on the charges of the particles and the medium.
\begin{figure}[httb]
 \centering
  \includegraphics[width=10cm]{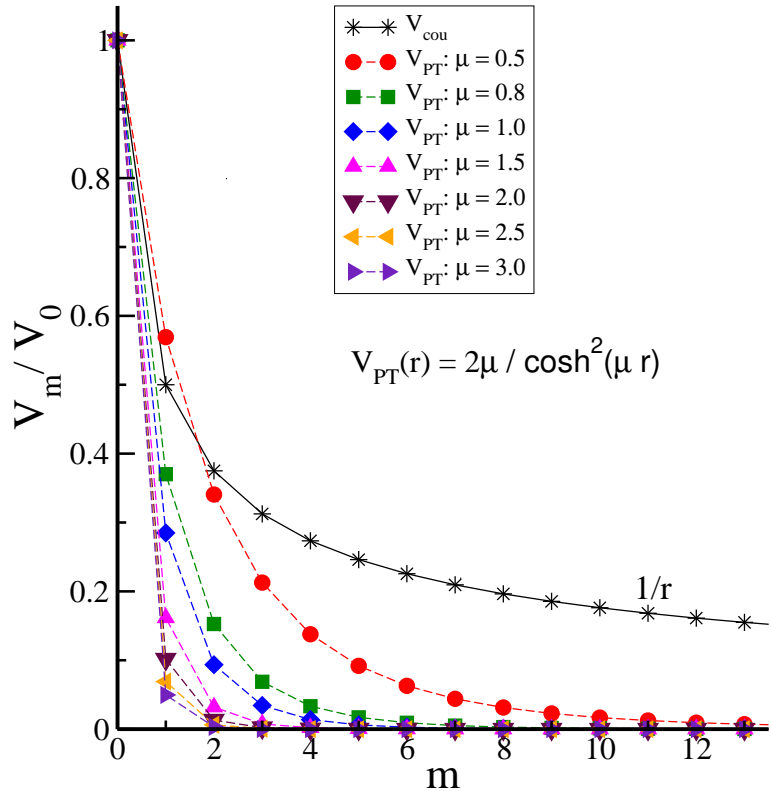}  
   \caption{Haldane pseudo-potentials for various range of interaction (variation of $\mu$) is shown by dotted lines. We only have $V_0$ relevant, for a higher values of $\mu$, or short-range $\delta$-type interaction. Large values of $V_m$ has a non-zero possibility for long-range interaction only i.e. small value of $\mu$. The similar figure using Coulomb interaction, which stays finite for large values of $m$, is shown by the solid black-line with an asterisk (*) [ Data taken from Ref. \cite{SDE_MI} ].}
   \label{PseV}
\end{figure}
\subsection{Haldane pseudo-potential}
The number that quantifies the interaction energy of a pair of particles is the pair pseudo-potential, which was introduced by Haldane \cite{Girvin} and is referred to the Haldane pseudo-potential \cite{pseudoV}. The pseudo-potential for the particles in $n-th$-LL on an infinite plane is written as $V_{n,m}$ or $V_m$; where, $m$ represents the z-component of relative angular momentum for a pair of particles. One can study the energy-spectrum and the whole dynamics of a system of $N$-particles residing in $n-th$-LL using the set of pair-energies provided by $V m$. In contrast to the typical (Coulomb) two-body interaction, Simon {\it et. al.} explored the counterpart of this pseudo-potential construction \cite{Simon2007} that emerges from generic multi-particle interactions.
The pseudo-potentials for the two categories of interacting potentials mentioned above are indicated below. As a function of the relative angular momentum $m$ of two particles, the pseudo-potential parameters $V_m/V_0$ are displayed in Fig. \ref{PseV}. Higher values of $m$ cause the pseudo-potentials $V_m$ for short-range Coulomb contact to go to zero, while they keep being finite for long-range Coulomb interaction. The most significant fact is that the Bose system only considers even values, whereas the fermionic system only considers odd values of $m$.
%
\section{Composite fermionization of Bose particles}
In dilute rotating confined atomic Bose gases, correlated states resulting from inter-particle interactions can be characterised by weakly-interacting CFs of Bose particles, such as bound states of an odd number of flux quanta  (say $f = 1,3,5, \cdots$) and Bose atoms \cite{CF_on_BEC}. Fermionic statistics apply to this bound state, which consists of a boson and an odd number of flux quanta, which are similar to vortices in terms of their topological characteristics. Although there exist an fictitious magnetic field $B$; a smaller magnetic field is felt by the Bose CFs,
\begin{eqnarray}
B^* = B - f \rho \phi_0 \;,
\end{eqnarray}
 where $\phi_0$ is the magnetic flux-quantum, and $\rho$ be the number density of the bose particles. In this weaker magnetic field, CF creates LLs, also known as $\Lambda$-levels. The relationship between the filling fraction of Bose CF (say, $n$, an integer number), and the LL filling fraction of the Bose atoms is provided by,
\begin{eqnarray}
\nu = \frac{n}{n f + 1} \;.
\label{CF_seq1}
\end{eqnarray}
This above mechanism demonstrates that, we can treat the interacting Bose atoms as fermions \cite{CP_statistics} and those filling fractions perfectly match the Jain sequence \cite{Jain_CF}.
\begin{figure}[h]
\centering
 \begin{minipage}{0.5\textwidth}
   \centering
   \includegraphics[width=8.0cm]{Eg.eps}
   \caption{The ground-state energy per particle is represented as a function of the interaction parameter $\mu$ in the unit of $0.01 U$ for the filling fractions $\nu = 1/2, 1/4,$ and $ 1/6$.  [Data taken from Ref. \cite{SDE_MI} ]. Ground-state energy is monotonically decreasing with increase of $\mu$-values.}
   \label{Eg}
 \end{minipage}
 \begin{minipage}{0.45\textwidth}
  \centering
  \includegraphics[width = 7.25cm]{pc.eps}
  \caption{pair-correlation functions are shown for those three filling fractions of bosonic FQHS [Data taken from Ref. \cite{MI23}].}
 \end{minipage}
\end{figure} 
\section{Ground state properties}
In literature, we noticed that for small enough filling fraction of rotating BEC, one may have formation of FQHS \cite{PRA77_2008}. People focused to study neutral collective excitation spectra of FQHE in an ultra-dilute rotating Bose system using CF theory. The composite fermionization of Bose atoms deteriorates \cite{Chang} with increasing the value of $n$ along the CF sequence (equation (\ref{CF_seq1})). Majumder and others accounted only first series of Jain's sequence with odd numbers of flux attachments to study the excitation spectra in this system. Moumita {\it et. al.} \cite{MI23} calculated the pair-correlation function of the first three fractions to confirm their strongly interacting liquid-like nature. They also investigated their ground state energy by varying the range of PT-interaction potential. People considered the standard spherical geometry \cite{Hierchi_1} in Monte-Carlo simulation, assuming $N$ correlated Bose-atoms residing on the 2D spherical surface of radius $R$ \cite{Haldane}. The ground-state wave function for the $N$-particles at filling fraction $\nu$, (i.e. $n$-number of filled $\Lambda$-levels of Bose CFs) is given by \cite{2Comp_BEC}
\begin{equation}
  \Psi_g = J^{-1} P_{LLL} J^2 \; \Phi_1 \left ( \Omega_1, \Omega_2, \cdots \Omega_N \right ) \;.
\end{equation}
Where the position of Bose CFs on the spherical surface is represented by $\Omega_i$, the Slater determinant (SD) of fully filled lowest $\Lambda$-level of Bose CFs is denoted by $\Phi_1$, and $P_{LLL}$ be the projection operator onto the LLL \cite{PLL}. To involve the interaction part another factor (Jastrow factor) is to be multiplied with SD; which is given by
\begin{equation}
  J = \prod _{i<j}^N \left (u_i v_j - u_j v_i \right )^p \;. \nonumber
\end{equation}
Following the fermionic transformation, the system's overall wave-function retains symmetric since $\Phi_1$ and $J$ both remain odd if we have interchanged two particles.\\
%
The ground state energy ($E_g$) can be calculated using the variational principle 
\begin{equation}
  E_g = \frac{<\Psi_g| V_{PT} |\Psi_g >}{<\Psi_g | \Psi_g >} \;.
  \label{E_GS}
\end{equation}
In Fig. \ref{Eg} variation of ground-state energies ($E_g$) for a different filling fraction as a function of $\mu$ have been shown for the PT interaction potential ($V_{PT}$), since it has the freedom to control the range of interaction. The energy reduces with the increase of the interaction parameter $\mu$, as the range of interaction reduces with $\mu$. The pair-correlation function $g(r_1 , r_2 )$ is the joint probability of finding a particle at the position $r_1$ with another particle located at the position $r_2$ \cite{PC}. For $N$-particle system it can be calculated as a function of relative coordinates between two particles;
{\small \begin{eqnarray}
  g((r_1 , r_2) = \frac{N(N-1)}{\rho } \int d^{2}r_{3}d^{2}r_{4} \cdots d^{2}r_{N} \mid\Psi_g (r_{1}, r_{2}, \cdots,r_{N})\mid^{2} \nonumber
\end{eqnarray}} 
It is an important physical quantity for liquids. It becomes to a constant value after a finite separation distance between two particles in case of liquid and oscillates in case of crystalline phase of system \cite{PC1}. 
\section{Neutral collective excitation of rotating Bose system} 
An essential objective of quantum many-body theory is the investigation of elementary excitations. There are neutral collective excitations of the quantum Hall (QH) fluid in which the density and charge ripples are in wave-like behaviour over large distances. The fact that these modes are gapped for the FQHS at wavevector $k = 0$, which represents the incompressibility of the QH liquid. The energy dispersion relation exhibits some minima depending upon the filling fraction at some finite wavevector $k$, referred to as roton-minima, similar to the collective excitations of the superfluid. There are two varieties of neutral collective modes: spin-conserving excitations (SCE), and spin-flip excitations (SFE). Here, we have focused to review the collective SCE and SFE for the bosonic FQHS in the thermodynamic limit ($N \rightarrow 0$). 
 \begin{figure}[httb]
 \centering
   \includegraphics[width=10cm]{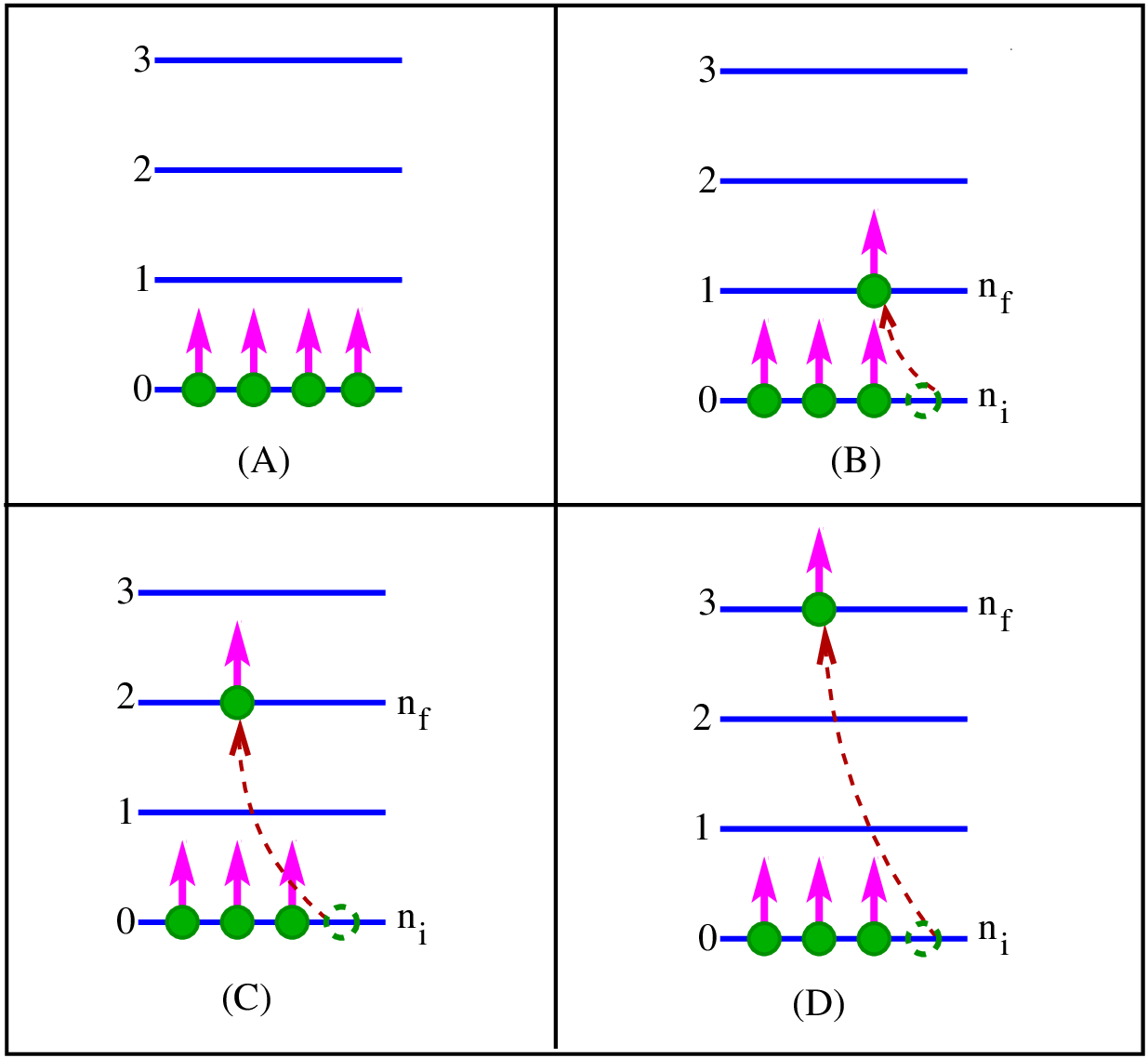}
   \caption{Schematic diagram of collective SCE: (A) The very first block shows the ground state for $1/2$ filling fraction. Rest of the three blocks represents three lowest order excitations. Block (B) represents the transition of Bose CF from $0$-th $\Lambda$-level to $1$-st $\Lambda$-level, block (C) denotes the transition of Bose CF from $0$-th $\Lambda$-level to $2$-nd $\Lambda$-level, and block (D) represents the transition of Bose CF from $0$-th $\Lambda$-level to $3$-rd $\Lambda$-level; leaving a hole in the $0$-th $\Lambda$-level. Here the $\Lambda$-levels are represented by blue straight lines, solid-circular dots are Bose atoms and arrow lines attached with them are the flux quanta and the dotted-circular dots in lowest $\Lambda$-level denotes the missing of Bose-CF i.e. hole.}
   \label{SCE}
 \end{figure}
\subsection{Spin-conserving excitation}
For a single-component fully-polarized condensate, one can study the excitation spectra considering CFs excite to a higher $\Lambda$-level. The phenomenology we've been talking about is therefore comparable to the QH effect of spin-less (fully polarised) electrons, i.e., the situation in which the effective Zeeman gap is so large that the electrons' spin degree of freedom is frozen out. The single-spin state for the bosonic system is named as spin-up $\uparrow$, although bosons has integral spin. So, when a Bose-CF changes from a filled $0$-th $\Lambda$-level to an empty $n_f\;$-th $\Lambda$ level of same spin-state, then the spin-conserving (SC) excited state wave function for $N$-particle system at that filling fraction $\nu$ is obtained from \cite{DM_SSM, Jain_SSM}, which is as follows,
\begin{eqnarray}
  \Psi_{ex}(L) &=& J^{-1} P_{LLL} J^2 \sum_{m_h} |m_h> <q, m_h; n_f+q,m_p|L,M> \;.
\end{eqnarray}
Here $|m_h>$ denotes Slater determinant of $N-1$ number of Bose-CFs lying in the lowest $\Lambda$ level and one Bose-CF in the $n_f$-th $\Lambda$-level with same spin having angular momenta $m_p$. The term $<q, m_h;n_f+q,m_p|L,M>$ is the usual Clebsch-Gordan coefficients, where $L$ is the total angular momentum and Z-component of that is taken as zero ($M=0$), to reduce the computation time without losing generality.

The superposition of all probable excitons incorporates actual collective excitation rather than a single CF-exciton state. Here three most-probable exciton states are taken into account 0$\uparrow$ $\rightarrow$ 1$\uparrow$; 0$\uparrow$ $ \rightarrow$ 2$\uparrow$ and 0$\uparrow$ $ \rightarrow  $ 3$\uparrow$ i.e. the SCEs occur if one CF jumps from 0$\uparrow$ to the other spin-up  higher $\Lambda$ levels.
\subsection{Spin-flip excitation}
We have so far discussed about spinless Bose-condensates or single-component system only. Finally, let's briefly discuss the more general case, what if the internal degrees of freedom, such as spin, has an important role. If the Zeeman energy of two hyperfine spin-states are lower than the cyclotron energy then transition can occur between two spin-reversed states too. Let us end with a brief discussion of the more general case where internal degrees of freedom, such as spin, play a role. Although the Bose atoms has integral number of spin ($0,1,2,..$), we named two hyperfines states of Bose-condensate as spin-up $\uparrow$ and down-state $\downarrow$ (similar to fermionic system).
\begin{figure}[httb]
 \centering
   \includegraphics[width=10cm]{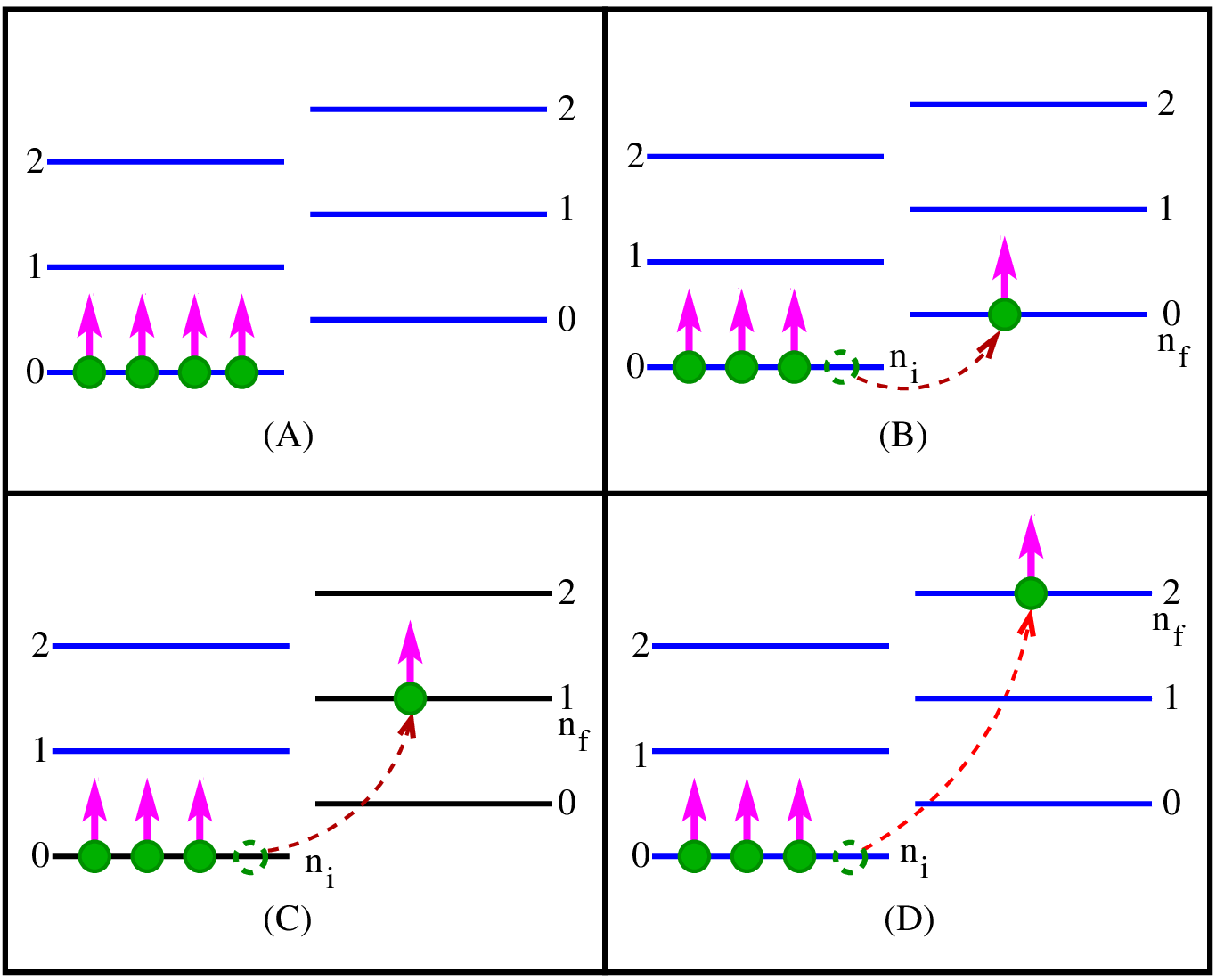}
   \caption{Schematic diagram of collective SFE : First block (A) represents the ground state of $1/2$ filling fraction since there are only one flux quanta attached. In each of the four blocks two-spin state $\Lambda$-levels are shown; up-spin $\Lambda$-levels are in the left side and down-spin $\Lambda$-levels are in the right side in each of the panels. Block (B) represents transition of Bose-CF from $0 \uparrow$-th $\Lambda$-level to $0 \downarrow$  $\Lambda$-level.  Block (C) represents transition of Bose-CF from $0 \uparrow$-th $\Lambda$-level to $1 \downarrow$  $\Lambda$-level.  Block (D) represents transition of Bose-CF from $0 \uparrow$-th $\Lambda$-level to $2 \downarrow$  $\Lambda$-level. All of these three excitations leaving behind a hole (missing of Bose-CF) in the $0 \uparrow$  $\Lambda$-level [Reprinted with permission from Ref. \cite{SDE_MI}].}
   \label{SFE}
 \end{figure}
The excited-state wave function for a spin-flip $N$-particle system, which corresponds to the transition of a Bose-CF from the filled $0$-th $\Lambda$-level to the other spin $n_f\;$-th $\Lambda$ level, is obtained similarly by taking the Slater determinants of  $N-1$ number of particles ($|m_h>_{\small{N-1}}$) in one spin ($\uparrow$) state along with one single-particle state ($Y_{Q,n_f,m_p} $) in another spin ($\downarrow$) state. The wave functions for SFE is written as,
\begin{eqnarray}
  \Psi_{ex}(L) &=& J^{-1} P_{LLL} J^2 \sum_{m_h} |m_h>_{\small{N-1}}\; Y_{Q,n_f,m_p} <q, m_h; n_f+q,m_p|L,M> \;.
\end{eqnarray}
The spin-flip excitations can be recognized as one Bose-CF jumps from 0$\uparrow$ to any one of the first three spin-down $\Lambda$-levels. Generally 0$\uparrow$ $\rightarrow$ 0$\downarrow$ transition is known as the conventional spin-wave (SW) excitation, whereas 0$\uparrow$ $\rightarrow$ 1$\downarrow$ and 0$\uparrow$ $\rightarrow$ 2$\downarrow$ are called the SFE. Possible collective SCE and SFE for $\nu = 1/2$ filling fraction are depicted in Figs. Fig. \ref{SCE} and \ref{SFE}. Similar excitations are taken into account for the other filling fractions. The Gram-Schmidt Orthonormalization method \cite{GS_ortho} was used to orthogonalize low energy exciton states with a fixed angular momentum because the excitons are not orthogonal.
\begin{figure}[httb]
 \centering
  \includegraphics[width = 12cm]{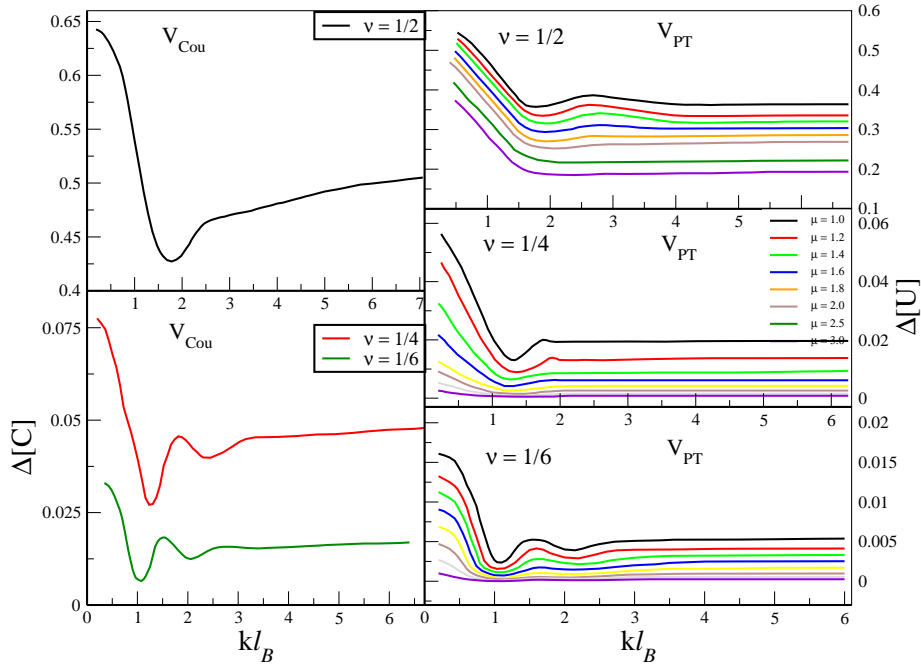}
  \caption{Collective neutral SCE: Lowest order energy spectra for different values of $\mu$ are shown in the three right-most boxes. $\mu$ values are increasing downwards in each filling fraction. Energy spectra using Coulomb interaction potential are shown also in two left-most boxes. Wave vector ($k$) is related to the total angular momentum by $kl_B = L/R$ [ Data taken from Ref. \cite{MI23, Physica_B}].}
  \label{spec_SCE}
 \end{figure}
 \begin{figure}[httb]
 \centering
   {\includegraphics[width = 12cm]{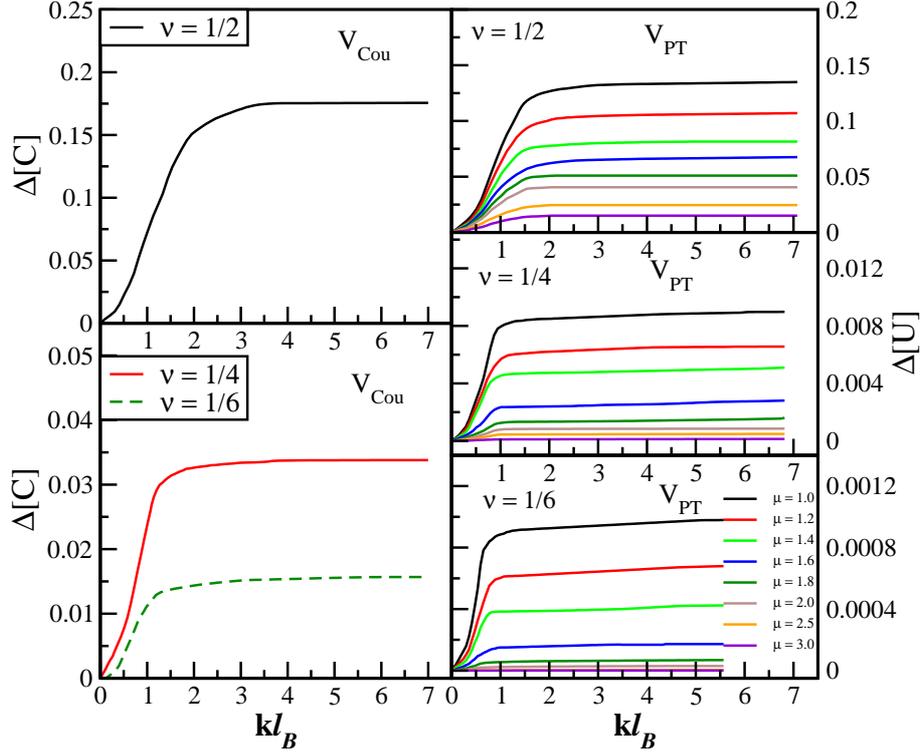}}
   \caption{Collective SFE spectra: The three right-most boxes display the lowest order energy spectra for the three filling fractions $\nu = 1/2, 1/4,$ and $1/6$ for various values of $\mu$. The value of $\mu$ are decreasing in upward directions in each of these three panels. In the two left-most boxes, energy spectra using the Coulomb interaction potential are also displayed. The equation $kl_B = L/R$ connects the wave vector ($k$) to the total angular momentum ($L$). [ Data taken from Ref. \cite{SDE_MI} ].}  
   \label{spec_SFE} 
 \end{figure}
The excitation spectra can be calculated as
\begin{equation}
  \Delta(L) = \frac{<\Psi_{ex}(L)| H |\Psi_{ex}(L)>}{<\Psi_{ex}(L) |\Psi_{ex}(L)>} - \frac{<\Psi_g| H |\Psi_g >}{<\Psi_g | \Psi_g >} \;.
  \label{spectra_Eq}
\end{equation}
As the system is ultra-dilute and rapidly rotating, we assumed that all the atoms are restricted in the LLL. Therefore, the Hamiltonian ($H$) of the equation (\ref{spectra_Eq}) is the interaction potential ($V$) between the Bose atoms. The bra-ket ($< >$) notation in the above equation (\ref{spectra_Eq}) basically denotes multidimensional integration which are computed numerically by quantum Monte-Carlo method. 
\section{Discussions} 
Jain and others \cite{Chang, 2018} tested the hypothesis that the interacting bosons map onto weakly interacting fermions carrying odd number of flux quanta each, which was inspired by the composite fermion theory of a the FQHE of electrons. There are some exact diagonalization calculations and CF calculations for a small number of particles \cite{exact_12}, which demonstrates that the excitation spectrum for FQHS of bosonic system at filling fraction $\nu = 1/2$ is not associated with the roton type of excitation \cite{noRoton}, whereas almost all the filling fractions of electronic FQHE contain roton minimum in their energy spectrum. So Majumder group first tried to check the existence of roton-minima for $\nu=1/2$ fractions only using PT interaction potential. They calculated the energy spectra for different numbers of particles from $80$ to $160$ for each of the three filling fractions $\nu=1/2$ \cite{Physica_B} in rotating BEC. Later they studied the same for the other two fractions $\nu=1/4, 1/6$ also \cite{MI23}. We have shown the average energy spectra for spin-conserving excitation in Fig. \ref{spec_SCE} and SFE in Fig. \ref{spec_SFE}. This clearly shows that the nature of excitation is similar to the electronic FQHE. At a low value of $\mu$, i.e. for long-range interaction, one can have a very sharp roton minimum, as we increase the value of $\mu$ i.e. decrease the range of interaction the roton minimum becomes shallow and vanishes after some limiting range of interaction. The energy is slightly higher than the energy spectra of delta-function interaction. One could not able to go as close as delta-function interaction as the ground state energy becomes very small. They observed that as one is going to the short-ranged region, the energy is reducing and it is expected that the energy will be identical with the delta-function interaction energy spectra. However, for the usual Coulomb interaction, there are sharp roton minima in the energy spectra, which confirms the disappearance of the roton minima is due to the short-ranged contact interaction between the Bose atoms.
Moumita {\it et. al.} \cite{SDE_MI} also studied the collective excitations with the spin degrees of freedom for this bosonic system. The low energy excitation is the spin-wave (SW) excitation. In the small wave vector limit ($k \rightarrow 0$), Larmor’s theorem stipulates that the SW energy is precisely equal to the bare Zeeman energy, for a conventional ferromagnet, such as the one at $\nu=1$ \cite{RPA_nu1} or $\nu=1/3$ \cite{SW}, SW has positive dispersion with energy that increases monotonically with wave vector reaching a large wave vector asymptotic limit of particle and hole separation energy with opposite spin. In their study, the SW excitation is also present in this Bose system. For spin-flip excitations, no roton minima are observed for those three filling fractions. The nature of the excitations is the same for both the two types of potentials, long-ranged Coulomb interaction and also for short-ranged delta-type potential. As one increase the value of $\mu$ i.e. decrease the range of interaction, the value of $\Delta$ becomes lower and vanishes after some limiting range of interaction and the spectra become flat. The interesting fact is that the nature of spectra does not depend on the range of interaction. From Fig. \ref{spec_SFE}, it is obvious that the energy should be reduced at a shorter range of interaction. So we conclude that the spin-flip excitation of those three filling fractions supports the SW excitation for all ranges of interactions.
\section{Future outlook}
An overview study on QH-physics in diluted rotating Bose atoms must be preliminary, since the research area is still quite active. It is pretty much true to say at this point that the theoretical aspect has been investigated thoroughly. The major practical hurdles to physically entering the QH-domain is the requirement that rotation speeds for the typical harmonic confinement be quite near to the deconfinement limit. Modifying the confining potential is an apparent technique to make rotation of the Bose-cloud faster than the strength of harmonic potential without it bursting apart.
Several theoreticians suggested to add a small quadratic-term to the trap potential \cite{chevy} like an anharmonic oscillator. Also some proposed to involve co-rotating optical lattices \cite{Olattice}, also non-rotating optical lattices \cite{Palmer}, with laser induced hopping \cite{Jaksch}. Using all these most recent concepts, new experimental advancements are undoubtedly possible in the near future.
\section*{Acknowledgement} One of the authors, Moumita wants to thank the institute post-doctoral fellowship grant (IIT Bombay) for her financial help.
\section*{Data Availability Statement} No Data associated in the manuscript.

\end{document}